# Interacting Ions in Biophysics: Real is not Ideal

Bob Eisenberg

Department of Molecular Biophysics Rush University

Chicago IL 60612

USA

and

Department of Chemistry and Miller Institute

University of California

Berkeley CA 94720

USA

May 1, 2013

Running Title: **Ions in Solutions Interact**

File name is "Interacting Ions Real is not Ideal FINAL May 1-1 2013.docx"

# **Abstract**

Ions in water are important throughout biology, from molecules to organs. Classically, ions in water were treated as ideal noninteracting particles in a perfect gas. Excess free energy of each ion was zero. Mathematics was not available to deal consistently with flows, or interactions with other ions or boundaries. Non-classical approaches are needed because ions in biological conditions flow and interact. The concentration gradient of one ion can drive the flow of another, even in a bulk solution. A variational multiscale approach is needed to deal with interactions and flow. The recently developed energetic variational approach to dissipative systems allows mathematically consistent treatment of the bio-ions $Na^+$, $K^+$, $Ca^{2+}$ and $Cl^-$ as they interact and flow. Interactions produce large excess free energy that dominate the properties of the high concentration of ions in and near protein active sites, ion channels, and nucleic acids: the number density of ions is often $> 10$ M. Ions in such crowded quarters interact strongly with each other as well as with the surrounding protein. Non-ideal behavior found in many experiments has classically been ascribed to allosteric interactions mediated by the protein and its conformation changes. The ion-ion interactions present in crowded solutions—independent of conformation changes of the protein—are likely to change the interpretation of many allosteric phenomena. Computation of all atoms is a popular alternative to the multiscale approach. Such computations involve formidable challenges. Biological systems exist on very different scales from atomic motion. Biological systems exist in ionic mixtures (like extracellular and intracellular solutions), and usually involve flow and trace concentrations of messenger ions (e.g., $10^{-7}$ M $Ca^{2+}$). Energetic variational methods can deal with these characteristic properties of biological systems while we await the maturation and calibration of all atom simulations of ionic mixtures and divalents.

Life occurs in ionic solutions. Pure water is lethal to most cells and biomolecules. The properties of most proteins depend on the details of the mixtures of ionic solutions found outside and inside cells. Trace concentrations ($<10^{-6}$ M) of $Ca^{2+}$ and other signaling molecules actually provide physiological control of many biological pathways and proteins inside cells.

Understanding the properties of living systems depends on the understanding of properties of ionic solutions: indeed, the early history of physical chemistry and physiology overlap remarkably probably for that reason. (Volta, Galvani and Fick were physiologists as much as physical chemists.) Biologists have, however, not kept up with advances in the understanding of ionic solutions, particularly ionic mixtures, understandably enough in my view, given how hard they have worked to provide the magnificent insights of structural and molecular biology.

Biophysicists are taught ideal equilibrium physical chemistry in which ions are points and flows are zero. Biophysicists have been taught this idealization for good reason: until recently the mathematics of interacting ions (of finite size) in flowing systems was not available. The mathematics of closely related (1) charge transport in semiconductors (2-6) could not be used—in its original form, despite initial enthusiasm (7,8)—because the charges that flow in semiconductors are points, with no diameter. Points cannot be crowded as bio-ions $Na^+$, $K^+$, $Ca^{2+}$ and $Cl^-$ often are in channels and active sites of enzymes (9). This situation has changed because of a recent development in mathematics.

Variational mathematics is now available to deal with ions of different diameters, interacting in solutions, as they flow. Historically, variational mathematics dealt with multiple types of forces in conservative systems (with Hamiltonians) that are difficult or impossible to describe with other methods. A generalization of this variational approach can now be used to describe systems that involve dissipative as well as conservative forces.

The purpose of this paper is to bring this energy variational approach to the attention of biophysicists, and to discuss the consequences for our classical understanding. Along the way, I point out the challenges that all atom simulations face as they try to deal with the realities of biological function.

**<u>Why do we need this mathematics?</u>** My biological colleagues at this stage often wonder why they need this new mathematics? Why can't they use the classical approach they learned in school? Many physical colleagues wonder what is different about this variational mathematics, why the big fuss?

The reason different mathematics is needed is that biological systems use the special properties of crowded spherical ions as they interact and flow in a tiny system say twice as wide as the ions themselves. Ions obviously interact electrically and by steric exclusion in such a system. Classical chemical approaches assume no interactions and no flow (i.e., equilibrium). There is no excess free energy from either non-ideal property. Classical mathematics of nonequilibrium electrical systems (like semiconductors) do not deal easily with crowded spheres of different diameter. The biology itself shows that the classical approaches of chemistry and semiconductors need to be extended.

**Excess free energy of the crowded ions cannot be ignored** because biology uses that excess free energy to perform its essential function of selectivity and perhaps other things as well. Evolution uses the properties of crowded ions to produce the selective properties that it needs in enzymes and channels. The ions are so crowded in and near enzymes (9) and channels (and nucleic acids) that their behavior is highly correlated. The classical references describing the properties of ions in non-ideal situations are (10,11) and excellent textbook treatments are (12-31). Reference (32) compares enzymes and channels. Other samples of the enormous literature are (33-47).

The new variational mathematics allows consistent treatment (with minimal adjustable parameters) of interacting spheres as they conduct and diffuse (48-55). It can easily be extended to deal with convection or heat flow in a mathematically consistent way, as it has in closely related problems (51,56) and problems of greater apparent complexity (57,58).

To summarize, the new variational mathematics are important in biology and biophysics because ions are crowded into a tiny space in and near protein (enzyme) active sites, ion channels, and nucleic acids. One can hardly imagine systems of greater biological importance. The crowding is dramatic, producing number densities often larger than 10 M (using classical chemical units of concentration, in which the number density of NaCl is ~37 M and liquid water is ~55 M).

**Crowded Ions as a biological adaptation.** These crowded conditions of ions are so special and so unusual that a sensible biologist would guess they are an 'adaptation'.

Evolution often uses unusual conditions or structures to perform life's functions. When a biologist finds an unusual condition or structure, it is usually productive to study how that condition allows the structure to perform biological functions. Study of such adaptations is useful, whether the study is of the panda's thumb or the flamingo's smile (59,60), the capillary of the squid that turned out to be a giant nerve fiber (61-63), or the twisted fabulously long sticky viscous polymer called DNA, eventually found to be the genetic code.

Unusual things in biology often turn out to be unusual for a reason. It is best to investigate and not ignore them.

**Enormous number densities of crowded charges** in and near ion channels and active sites are unusual, found in only a few other places I know of. Enormous number densities are also found in crucial locations in our electrochemical technology, suggesting a generalization, as a productive working hypothesis, for both electrochemistry and biophysics: I suggest that where ions are important they are likely to be highly concentrated (so 'conductances' and currents are large) or very depleted (so resistances are large and flows can be controlled). We note that many of the most important properties of semiconductor devices are produced by depletion zones (2-4,6,64-67).

Interactions are important in enzymes, ion channels, and transporters, because they can produce coupling of fluxes usually attributed to proteins, even in the absence of a protein. When interactions dominate, ions in bulk solution can flow uphill against their own gradient of activity (e.g., p. 377 of (68)). The energy for uphill flow of one ion comes from the downhill gradients of other ions.

Free energies of steric exclusion are significant in ordinary Ringer's solutions as well as in and near macromolecules: $Na^+$ and $K^+$ differ because they have different diameters. The different diameters of $Na^+$ and $K^+$ is why the activity of these solutions does not equal their number density as shown by a large literature of physical chemistry previously cited. Great attention has been paid to ionic interactions in chemical engineering (13,15,16,21,24,25,28,30,69-74) and geophysics (75-78). To oversimplify, $Na^+$ and $K^+$ are identical when they are ideal. They differ because they are non-ideal. The different roles of $Na^+$ and $K^+$ are essential for a wide range of biological function. $Na^+$ and $K^+$ cannot be treated as ideal ions in the context of biology. Theories and simulations must compute the (non-ideal) properties of $Na^+$ and $K^+$ with reasonable accuracy.

**Interactions define channels and transporters.** In classical physiology, the interactions of ionic flux have central importance. Indeed, Hodgkin used the interactions of ion fluxes to define channels and separate their properties from fluxes in bulk solution, on the one hand, and transporters (79,80), on the other, before channels were known to be proteins, before anyone had glimpsed the structure of any protein let alone a channel protein. Bass (81-84) provided a mathematical analysis based on Hodgkin's approach to channels. Hodgkin and Bass assumed independent behavior in the baths and neglected both steric interactions and the role of the charge of ions in creating the electric fields they move in. (A more modern, more realistic analysis might be helpful.)

In classical physiology, interactions in channels were treated by rate models of single file systems (85,86). These sadly did not compute electrostatics and do not deal with friction or thermal Brownian motion. Thus, it is hard to know what to make of the results, since electrostatic and frictional forces (87,88) dominate such small highly charged systems which move ceaselessly in thermal motion (88-90). Difficulties with biophysical rate theories have been extensively discussed in the biological (7,91-99) and physical (100-105) literature. Models of single filing that deal with electrostatics and friction consistently are just now emerging (106), as far as I know. The anomalous mole fraction effect was once thought to be a sure sign of single filing (85,86). We now know otherwise (107-110).

In transporters, enzymes and binding proteins, interactions were assumed to arise in the molecular mechanism of the protein(s) that make up the system. When a flux of $K^+$ was sensitive to $Na^+$, we thought of allosteric interactions as in enzymes. See the comparison of enzymes and channels in (32). The possibility of ions interacting themselves, electrically and from steric exclusion, was not considered, probably because the ions were treated classically, without excess free energy being included in the relevant equations.

Allostery plays a large role in classical and contemporary biochemistry and biophysics, as reviewed recently (111,112). Theories of allostery assume ideal properties of substrates and rarely include 'background' ions at all. Theories rarely if ever include excess free energy terms of the substrate.

If substrate and ions are crowded together in binding or active sites, it is hard to see how they could avoid interactions. It is hard to believe that all the interactions ascribed to allosteric properties of the protein are independent of the substrate-substrate, ion-ion, and ion-substrate interactions.

**Interpretations of allosteric phenomena will need to be reconsidered** with theories that allow all components to interact, in my opinion. It will be necessary to make an explicit model of each binding interaction, and conformation change, computing the free energy change of all components using a model that allows them all to interact. This daunting task has barely begun. One can expect general principles and simplifications to emerge only from analysis of many specific cases (as in much of biology (59,60)).

**Interactions in Ion Channels.** In one area of biology, studying the interactions of ions has been surprisingly successful. The selectivity of the calcium channel $Ca_V$, the sodium channel $Na_V$, and the ryanodine receptor RyR have been understood quite well using a primitive model of the channel structure and an implicit model of solvation, in the spirit of the implicit solvent primitive model of ionic solutions.

The primitive model of ionic solutions accounts for the fundamental property of ionic solutions—activity or chemical potential—over a wide range of conditions in a variety of solutions, as well reviewed by (17,18,28,47). The free energy per mole, or the excess free energy per mole, or the activity, or the activity coefficient of solutions are accounted for better than in many high resolution calculations of the properties of ionic solutions (20,21,30,47,54,113-118), although a great deal of work is going on to improve these higher resolution models (48,50,52-54,119-126). (I am purposely imprecise with the complex nomenclature and units of physical chemistry. Textbooks define these precisely (17,18) and a most useful set of standard symbols and definitions is available in the 'green book'(127)).

Nonner and Eisenberg introduced (128-132) primitive models of channels in which the protein is described by a few of its amino acid side chains confined to a tiny selectivity region ('filter') in the channel. An early review is (133). Solvation by water and by the channel protein are described by dielectric coefficients, as in the implicit solvent model of ionic solutions (24,134-139). These models were then studied with the Monte Carlo methods developed—and extensively tested—for physical systems by Boda and Henderson (140-143) in a series of more than thirty papers, reviewed in (98); see (125,144) for more recent references. The key papers describing the Ca$_V$ channel are (145,146). The key paper describing the Na$_V$ channel is (147), extended by (144). The ryanodine receptor is described in a series of papers led mostly by Dirk Gillespie (107,109,125,148-163) with key results in (158) and its supplementary material.

**Surprising success of simple models.** Nonner and Eisenberg were greatly surprised at the success of such simple representations. (Some details of the success are discussed below.) After all, these simple models omit most atomic and molecular details, but the utility, even necessity, of reduced models is now widely recognized, judging by their increasing use (164-172), even in simulations involving quantum and molecular mechanics.

The reduced models of the calcium channel account for most of the selectivity properties known in a wide variety of solutions of variable composition. These properties arise from strong interactions between ions and side chains. Rate constants and free energies computed from this model vary enormously (rate constants by more than a factor of 1,000) as conditions are changed (125,129,132,133,144,158,161,162,173-175).

**Success Depends on Computing Interactions.** Indeed, the model is successful because it computes the changes in interactions successfully. Simulations or theories that use a single value of rate constants or binding free energies do not allow changes in interactions with conditions. They do not even allow Debye-Hückel shielding (7,97,176) which is a general and unavoidable property of physical systems with mobile charge (177).

Ions and side chains of the protein are described in the successful reduced models by spheres with their 'crystal radii'. Radii are never changed as solutions are changed. Few parameters are needed to compute the binding curves of the channel over six orders of magnitude of concentration, just the effective diameter and dielectric coefficient of the protein and surrounding baths.

**Channels have been built using the primitive model.** Calcium selective channels can be constructed as suggested by the theory. OmpF porin, a hardly selective beta barrel bacterial channel that has no similarity to the calcium channels of eukaryotes, becomes calcium selective when glutamates are introduced by mutation (178) in a suitably narrowed space (178-180). Interactions produce the selectivity. Narrowing the space increases the interactions and the selectivity.

**One theory and one set of parameters describes very different channels.** If the side chains of the model are changed, the selectivity of the channel changes from that of an EEEE or EEEA calcium channel to that of a DEKA sodium channel, as found in experiments (181,182) discussed in detail in (98,147,175,183).

It is thus possible to account for the main selectivity properties of two of the most important voltage sensitive channels *with a single model without changing parameters*, just by changing side chains, as in experiments. It is important to note that the properties of DEKA and EEEA channels are very different and occur on different scales of concentration. I am unaware of any other models of selectivity in channels that can account for such dramatic changes in experimental properties without changing parameters.

**Different parameters change different properties**. Reduced models of this type use effective parameters, much as molecular dynamics uses effective force fields, that use macroscopic properties to estimate (and to represent) atomic scale forces. These parameters usually are composites and for that reason one expects changing a (composite effective) parameter like dielectric coefficient to change 'everything' that is observed and not to have a simple effect. One does not expect simple relations between selectivity and a single composite or effective parameter. And that is what usually happens. There is no (known) simple relation between parameters of the calcium channel models and selectivity. There is no simple relation between parameters of the sodium channel and selectivity between (for example) $Ca^{2+}$ and $Na^+$.

Simple relations are not found even in the most thorough analysis of selectivity in calcium or ryanodine channels. Boda's analysis (174) used Monte Carlo methods combined with Widom's insertion method. Gillespie's analysis (158) used Rosenfeld's density functional theory to determine components of free energy of binding. Both methods are 'state of the art' and did not produce a simple explanation of the selectivities examined. Neither method has yet been generally applied to other selectivity problems, probably because of the difficulty of implementing them properly.

The DEKA sodium channel, however, is different when we consider selectivity of $Na^+$ over $K^+$:

(1)   The selectivity of the DEKA sodium channel for $Na^+$ vs. $K^+$ in fact depends only on a structural parameter—the diameter of the channel—and not on a meaure of solvation—the dielectric coefficient. In the primitive model, solvation appears as a dielectric coefficient. See Fig. 1, redrawn from Fig. 8 of (147).

(2)   The contents of the DEKA channel depends only on the solvation (dielectric coefficient). See Fig.2, redrawn from Fig. 9 of (147).

(3)   The selectivity of the DEKA channel depends on the structure—the diameter of the channel. But selectivity does ***not*** depend on solvation (dielectric coefficient) and contents do ***not*** depend on structure (channel diameter).

(4)   Solvation and selectivity operate independently (to the amazement of the authors of (147)). Dielectric coefficient and diameter have separate effects. One parameter determines one thing and not the other. They are orthogonal.

The reduced model evidently captures the (free) energies used by this system to control selectivity and contents. These energies are controlled by simple reduced variables. It seems as if the reduced model has captured the adaptation used by evolution to control selectivity and conductance in this case.

**Gillespie's model of the ryanodine receptor** provides another example in which the reduced model can compute the non-ideal interactions that determine selectivity and (in this case) conduction. Here, Gillespie has been able to calculate current voltage curves in more than one hundred solutions—(158): many results are in the supplementary material—predicting subtle mole fraction effects before they were measured (107,109,125,158,163). Most strikingly, mutations involving drastic changes in the density of permanent charge (from some 17 molar to zero) are accounted for in several solutions.

**Interactions can be calculated in reduced models with realistic complex properties.** These successes show that computation of interactions is feasible in some biological situations and lend hope that similar approaches may be successful in the future in dealing with other biological systems that have selectivity and crowded charge, like active sites of enzymes.

Given the success of this work, one may wonder why new mathematics is needed. The answer is that a general method of dealing with interactions can be extended to situations not accessible to the Monte Carlo simulations used for calcium and sodium channels, and hold significant advantages over the methods used in dealing with the ryanodine receptor (details in (97-99,176)).

**A different approach: the transistor tradition**. The work described above arose in the chemical tradition, emphasizing the thermodynamic properties of systems at equilibrium, doing statistical mechanics in the thermodynamic limit, where boundary conditions are not involved. The theory of simple fluids is a magnificent example showing what the classical tradition can do when exploiting the simplifications produced by the thermodynamic limit (184-187).

Quite a different tradition is used to analyze systems that depend on flow and we turn to that tradition now in the context of semiconductor physics. The new variational mathematics will be used to unify the two approaches later on. There, we will view ionic solutions as complex fluids (48,50,51,54,122), and advocate (97,176,188) the use of mathematical methods designed to deal with the flows and interactions of complex fluids.

**Transistors are a system of charge transport closely related to ion channels** (1). Charge transport in transistors allows most of our modern technology. Integrated circuits and thus our digital technology depend on the flow of charged quasi-particles, holes and 'electrons'. The electrons of semiconductors are quasi-particles that I like to call 'semi-electrons'. Quasi-particles have only a faint resemblance to the isolated electrons of high school physics, despite their identical name.

Quasi-particles are most useful mathematical constructions that help us understand current flow in semiconductors because quasi-particles follow laws much like those an ion would follow if it were a point. But the quasi-particles—holes and semi-electrons—do not exist outside of semiconductors like silicon and germanium. Bio-ions exist permanently.

Bio-ions do not recombine, although of course weak acids (e.g., glutamic acid) and bases (e.g., lysine) do, including acidic and basic side chains of channel proteins. I have speculated that such recombination may be important in the function of transporters (7,8). Protonation of side chains would change the electric field and thus might create or control the correlated gates responsible for the ping-pong, alternating access mechanisms of transporters. I hasten to say this remains an idea, not a worked out model, let alone a fact.

Field effect transistors *FETs* contain channels in which the flow of quasi-particles follow the drift diffusion equations (4,6,64,189-191) with forces calculated from all the charges present, using Poisson's equation of the electric field. Quantum mechanics creates the underlying properties of semiconductors (and biological solutions, for that matter), but quantum mechanics enters indirectly in the classical theory of semiconductors (2-4,6,64,66,67,192). It determines the band structures and the properties of holes and semi-electrons. Direct computation of quantum effects are not needed. The drift diffusion equations are enough to deal with most properties of interest (4-6,193-195).

**Quasi-particles and real ions flow under the influence of electrical forces created by their own charge,** in large measure**.** Poisson or Maxwell's equations—that relate charge and electric forces—must therefore be solved along with the drift diffusion equations. Treatments that solve both equations together are called consistent. Treatments that do not, are called inconsistent. Classical Langevin equations of thermal motion are inconsistent for example when applied to ions in water because they assume a constant electrical field and do not compute it (196). Molecular dynamics simulations of thermal motion on the other hand are consistent, if the electric field is computed correctly so charges and electric potentials are related by Maxwell's equations.

Drift diffusion equations are a part of a multiscale analysis. In a multiscale analysis, a different set of (high resolution) equations are used to describe atomic properties. The high resolution equations can be used to show the existence of intermediate scale properties—for example, holes and semi-electrons—but they rarely allow direct computation of the properties of devices. The equations that describe practical properties of devices are at an intermediate scale. The intermediate scale drift diffusion equations are used very widely to construct (197-199) and describe semiconductor devices (6), although of course higher resolution models are needed and used (2,3,66,67,192,200,201) in some situations.

***PNP* and Drift Diffusion.** The drift diffusion equations are called *PNP* in biophysics to emphasize the importance of computing the variable spatial distribution of potential from the much less variable distribution of fixed ('permanent') charge, using the Poisson equation with boundary conditions (for bath concentration and potential), as extensively discussed in (7,8,52,54,95,106,120,123,144,148,202-253).

The name *PNP* for Poisson Nernst Planck (254) was introduced deliberately as a pun in a Biophysical Society Workshop (255) to emphasize (a) the analogy between ions in channels, and quasi-particles in transistors; and (b) the importance of computing the electric field, as opposed to assuming that the electric field is constant (256-258) in space, or as conditions, concentrations, or solutions change.

The electric field of *PNP*—like the electric field in transistors—is not constant as conditions change. Electric forces must be computed from all the charges present, including the ions and quasi-particles, as a consistent mathematical solution of the entire system. Most previous work on Nernst-Planck equations in biology and chemistry (for example, (91,230,241,259-262)) (a) did not mention the analogy with transistors (however, see (263-265)); (b) did not mention the importance of permanent charge (i.e., 'doping'); and most importantly (c) did not mention the crucial role of the **variable** shape (i.e., 'conformation') of the electric field and its large changes when bath concentrations or potential is changed.

The title of the early paper *'Computing the Field'* (7) was chosen to contrast with earlier approaches that used more or less constant fields (85,256-258), or unchanging potentials of mean force (266,267), free energy barriers (85,267-269), and rate constants (85,268,269). Fields describing forces must change as conditions are changed in consistent models and so cannot be assumed to be constant in shape, let alone constant in space. Fields describing forces are outputs of a consistent theory or simulation. Fields must be computed. Electric fields in channels and proteins cannot be assumed to be constant (7) if the electrical potentials in the system are to be consistent with the charges in the system.

**Transistors function by changing the conformation of the electric field** produced by doping and boundary conditions. The change in shape of the electric field is crucial for the function of transistors. Drift diffusion without doping, Poisson, or variable shapes of electric fields has a limited range of behaviors. With doping, Poisson, and variable shapes of fields, *PNP* can do everything a transistor and thus everything a computer can do. For example, elementary texts show how a single *FET* can be an amplifier, limiter, switch, multiplier, logarithm or exponentiator (6,65,270,271). Arrays of *FET*s provide all the logic, memory, and display functions of a computer. Solutions of the *PNP* equations can do everything a computer can do!

**Evolution needs devices as much as engineers do.** It seems unlikely that evolution would entirely ignore the devices that (ionic) *PNP* equations allow. It seems likely that evolution uses electric (and steric) fields that change shape to help with the function of proteins, channels, transporters, and enzymes (8). It seemed (7,8)—and seems (106)—possible that some functions of proteins customarily attributed to changes in the conformation of mass might actually be produced by changes in the conformation of their electric (and steric) fields. Transistors function by changing the conformation of their electric field without changing the conformation of their masses.

***PNP* equations are not enough** because the diameter of ions has important effects. The finite diameter of ions introduces correlations not found in the *PNP* equations. The *PNP* equations are just early members (i.e., low order terms) of an hierarchy of equations (272-278) like the BBGKY hierarchy (184,186,187) of equilibrium statistical mechanics. The correlations are important in ionic solutions and biology as we have discussed at length. Correlations produce nonideal behavior.

**Ionic Solutions are not dilute gases.** It seems clear that classical models of ionic solutions taught to biophysicists need to be replaced. Ionic solutions have interactions not found in uncharged, noninteracting ideal gases (279). Classical models of ionic solutions are poor models for that reason.

A poor model of this sort cannot explain interactions seen in experiments. If interactions are found experimentally, classical models will attribute them (mysteriously) to a single file in a channel protein, to the channel protein itself, to the transporter, the enzyme, or the nucleic acid. The explanations will be vague because they will not fit measurements over a range of conditions. Classical models have no other way to explain interactions. They will necessarily attribute interactions to the protein or to mysterious 'chemical effects' and rate constants because classical models do not consistently calculate interactions between ions or between ions and side chains of proteins.

**What model should we use to deal with interactions?** It seems clear from the previous presentation that the primitive model does well with some types of ion channels, and should be used as the initial approximation in similar cases, in which side chains of the channel protein mix with ions in the pore of the channel.

What is not so clear is what model should be used for bulk solutions (or for other biological systems, for example). The unfortunate reality is that physical chemists today have not yet found a good model for pure solutions with ionic strengths greater than 100 mM, like those found in biology, let alone for solutions of divalents, or for solutions like the ionic mixtures (of cytoplasm and 'Ringer's solutions') in which almost all life occurs.

**Understanding of bulk solutions is limited to special cases.** A recent summary paper (280) of a definitive book (25) summarizes present knowledge of specific ion effects. It says (p.11) "It is still a fact that over the last decades, it was easier to fly to the moon than to describe the free energy of even the simplest salt solutions beyond a concentration of 0.1M or so."

These feelings of contemporary physical chemists are not very different from those of the 1950's. Then, the standard textbook (12) of Robinson and Stokes—still widely used and in print—said, when talking about ionic solutions like seawater or Ringer solutions, "… many workers adopt a counsel of despair, confining their interest to concentrations below about 0.02 M, ... " (p. 302 of reference (12)). Note that very little biology occurs in solutions with concentrations below 0.02 M.

These feelings of despair arise out of frustration, in my view. Nowadays, as in the 1950's, data is available but understanding is not. Tremendous compilations of experimental data have been published (10,12-15,21,24,25,28,281). Two (15,21) are the result of large governmental projects, yet many workers use huge look up tables (21) because of the inadequacy of phenomenological models and their inability to predict behavior in conditions different from those used to generate the models. The models do not produce 'transferable' results in the jargon of the trade. Results measured in one set of conditions do not transfer to other conditions, because interactions are different in the two cases.

The need for understanding is great, the understanding is little, so frustration and despair are the predictable result.

**Response to despair: atomic simulations.** The constructive response to despair in academic as in real life (often) is to move along and try something—anything—new and different.

Atomic simulations are something new. Most efforts, for several decades, have focused on simulating the properties of ionic solutions by computing the motions of individual atoms on a time scale of $10^{-15}$ sec. The fantastic improvement of semiconductor technology—by a factor of some 400 billion in 68 years—called Moore's Law (282,283) has made it possible to compute some $2^{(2013-1955)/1.5}$ faster and better in 2013 than in 1955 (284). It is only natural to hope that direct simulation will do better than the counsels of despair. (It is interesting that the technological triumph of transistors was catalyzed by the *PNP* equations. *PNP* allows optimal scaling of semiconductor properties and device characteristics, as devices are made smaller (197,198). *PNP* speeds technological development because it often replaces slow, expensive trial and error experimentation with direct computation.)

The problems in reaching the biological scales of function are much harder, however, than commonly realized (285). In biology, not only does one face extrapolations of $10^{10}$ in time scale and length scale, but one also faces severe problems in dealing with the trace concentrations of messenger molecules ($Ca^{2+}$, hormones) so important in biology. These problems must be solved all at once, in one calculation, because biology deals with these issues all at once.

**Problems of calibrating simulations** of ionic solutions are just now being faced in a (fortunately if belatedly) growing literature. Recent contributions include (22,37,47,118,167,286-295). As far as I know, no one has attempted to calibrate simulations of seawater or its close relatives, intracellular and Ringer's solutions.

Even the issue of calculating the electric field has not been faced forthrightly in molecular dynamics simulations of ions or proteins. Few papers actually show that the electric field estimated by molecular dynamics is in fact that which would be produced by all the charges in the system being simulated, using Maxwell's or Poisson's equation. Arguments abound, concrete calibrations are scanty.

Calibrations are frequent and automatic in simulations of computational electronics. Transistor simulations and codes are routinely calibrated and checked to be sure the potentials and charges are consistent, in my experience (201,229,296-301).

All these problems of molecular dynamics can be solved, I suspect, with decades of future work. Future work must deal explicitly with the unavoidable problems of calibrating atomic simulations of macroscopic systems. Avoiding problems rarely solves them.

**It is important to keep in touch with all atom simulations in semiconductors** (see (302) and references cited there and in (6) and at website (2)). It seems that progress is possible but slow. Calibrated simulations of the tiniest semiconductor systems and transistors are possible (303-305). It is important to try to extend such methods from crystalline semiconductors to more disordered fluid systems like ionic solutions.

**Different models at different scales.** The central issue in ionic systems is the need for different models at different scales. Direct computation cannot fill 10 orders of magnitude in time, and space, and concentration, all together, all at once. Interactions among ions make this task essentially impossible if each interaction is computed one by one (as in molecular dynamics simulations) because ions interact on the macroscopic scale of millimeters to meters as they produce physiological behaviors like an action potential, that depend in an essential way on the dielectric properties of membranes.

**Dielectric problems have more than pairwise interactions.** The electric field in the presence of dielectric boundaries (i.e., lipid membranes (306-308)) cannot be described by pairwise interactions, as a matter of simple mathematics. The induced charge at the boundary couples every particle to every other particle making the number of interactions beyond astronomical and jeopardizing even the usual statistical mechanical definition of the state of one atom. If the 'state' of one atom depends on the coordinates of a macroscopic number of other atoms, it is not clear that the idea of state (as usually used in statistical mechanics) is useful or well defined.

**Action Potentials link atomic and macroscales.** Biophysicists have known for more than thirty years (since Neher and Sakmann (309-311)—and suspected for more than 60 years (since the work of Hodgkin and Huxley (312) if not Cole (313-316)—that the action potential is produced by atomic scale changes in structure controlled by electrical potentials far away, millimeters or centimeters away (in squid axon). They have known that action potentials propagate over meters and depend on properties of small groups of atoms ('selectivity filters' and 'gates') of individual molecules of channel proteins. They have known that statistical properties of single channel molecules ('the open probability' for example) control macroscopic potentials and also depend on those potentials. Thus, biophysicists have known

'forever' that computations of action potentials (along with many other biological functions) must involve both atomic length scales and biological time and length scales. What they did not, and do not know is how to span the gaps between atomic and biological time scales.

In fact, it seems clear to me that there can be no general way to deal with these gaps in scales. I think it is easy to create counterexamples to any attempt at a general approach. One simply has to add a process that is invisible on one scale. Measurements in the invisible region can never detect the invisible process. But the invisible process can dominate on another scale. Think of an added linear resonance process for a trivial example in a linear system. Added nonlinear processes can obviously have even more dramatic effects.

**Guess and Check.** Instead of a general multiscale theory, it seems to me that one needs to guess simplifications and check them, hoping to find the simplifications that evolution has chosen to fill these enormous gaps. The simplifications imposed by evolution may make primitive models of biological systems more useful than primitive models of physical systems, as they seem to make primitive models of calcium channels more useful than primitive models of bulk calcium solutions.

**Atomic scale structures control most of biology.** We all know that atomic scale structures control most of biology. We all should know that these atomic scale structures move (102) more or less at the speed of sound (angstroms in $10^{-12}$ sec). We all know that most biology occurs on a millisecond to second time scale. What we do not usually know is how evolution spans these gaps, except in a few cases, like the propagating action potential just described, where the electric field spans the gap because the electric field exists on all scales. It is inherently multiscale.

**Multiscale analysis and the action potential.** Multiscale analysis is familiar to classical biophysicists but under a different name. The classical analysis of the action potential spans scales by using the cable (i.e., telegrapher's) equation (317).

Multiscale analysis of the action potential is dramatically simplified because most ionic channels function independently (85,311). The currents from separate channel molecules can be simply added to produce the macroscopic response. Markov models are not needed. Any complete summation of the currents through single channels—e.g., by nested convolutions of single channel records—will do fine, as long as the summations include all behaviors and openings.

**Biology requires explicit intermediate scale models.** Thus, I would argue in general that working out biophysical mechanisms will require explicit intermediate scale models linked together in an hierarchy starting with atomic structures, winding up with the biological function itself.

These models must of course include the substances known to flow and to modulate flow in the biological function. And here we are back to interactions and ions. Almost all properties of membrane proteins and nucleic acids are sensitive to the type and concentration of salts in their environment. Most enzymes behave differently when $K^+Cl^-$ in surrounding solutions is replaced by $Na^+Cl^-$. Most enzyme reactions have different rates when $Ca^{2+}$ is added to the surrounding solution or when concentrations of any of the bio-ions $Na^+$, $K^+$ or $Ca^{2+}$ are changed substantially. If rates of enzymatic reactions depend on concentrations, (free) energies depend on concentration. Thus, some of the energies of the enzymatic reaction depend on the concentration of interacting ions of non-ideal ionic solutions. The energies are not ideal. Excess free energy is significant.

**Incomplete models are unlikely to be useful.** If such excess (i.e., non-ideal) energies and interactions are not included in the description or models of enzymatic activity, the models are incomplete. In my opinion, the models are unlikely to allow understanding and control if they are so incomplete. Much biological control comes from trace concentrations of $Ca^{2+}$ and other messenger molecules perhaps concentrated a great deal in selective binding sites. It seems very likely indeed that evolution will use interacting ions to control and energize protein function. Part of the power supply of enzymes is likely to come from the enormous concentration of interacting ions nearby.

For these reasons, bio-ions and messenger molecules must be included in the hierarchy of models. Bio-ions and messenger molecules interact with each other through electric and steric fields. Ionic interactions, both steric and electrical, must be naturally included in the hierarchy of models.

Here the power, precision, and generality of the laws of electricity are an enormous help. Maxwell's equations describe the flow of current no matter what the chemical nature, or atomic or macroscopic (or nuclear or astronomical) scale of the current carrier. Where electrical phenomena are involved, it is often possible to ignore chemical and atomic detail. Appropriate approximations (like Kirchoff's current law and the telegrapher's equations of cable theory) can be used to link scales. The power and importance of Kirchoff's law can be appreciated when one realizes that it implies perfect correlation (within the accuracy of Maxwell's equations, something better than 1 part in $10^{18}$) of ion motions, in contrast to the perfect independence of motion of the ideal particles assumed in classical models.

**Models must include flow.** The hierarchy of models needed to link atomic structure and biological function must also include macroscopic flows and current because most biological function requires flow. Flows cease only at death.

Classical models and approaches do not include interactions, current, or flows. Fortunately, classical mathematics has been extended so we now know how to deal with interactions and flows without guessing.

Mathematics has recently been developed to do what Onsager (318-324) and Curran (325) and Katchalsky (326) tried to do so valiantly. It is now possible to write the free energy and dissipation (i.e., friction) of components of a system, combine them in a natural way, and then uniquely derive the mathematical description of the behavior of this interacting system.

This new energy variational method is a generalization of the variational methods long used by physicists to describe mechanical, friction-free systems. The breakthrough is the inclusion of friction and flow in a well-defined way consistent with remarkably successful work in computational and theoretical fluid dynamics (327-331). The methods of this energetic variational calculus allow a (nearly) unique set of predictions given a model of the free energy and friction of a system.

**Energetic Variational Methods** *EnVarA* are discussed in tutorial detail in (48) which extends to ionic solutions the earlier work of other authors on more complex systems (49,56,332-341). Variational methods of this sort (50-53,58) and others (119,120,122,138,250,342) have been applied to a range of problems involving ionic solutions. These methods have been used, in a somewhat less general form, in the work on liquid crystals (343,344) that help makes possible the *LCD* technology we use every day.

The energy variational methods are not magic, of course; the predictions are no better than the physical models themselves. If we use an inadequate physical model, we will get an inadequate result, but we cannot know what part of the model is inadequate until we correctly compute the model, under a range of conditions and compare those computations with experiments. The new mathematics allows one to compute models including friction and predict—by mathematics alone—what the model will give in a variety of conditions, including interactions and flows. Models can be tested and improved much more quickly if predictions can be made (with minimal adjustable parameters) and tested over a wide range of experimental conditions. Otherwise, comparing models tends to be an ill posed problem.

**Energetic Variational Approach.** The methods of the energetic variational calculus are a promising new approach. These methods allow natural and mathematically consistent treatment of interactions and of flows but they depend on models of ions in solutions and proteins and these are not general. In the case of some ion channels, exceedingly simple models work, but there is no reason to believe such success will apply to other systems until the models actually succeed. The problem is the models, not the methods: numerical procedures (52,120,122,250,345) are now available to allow computations of interacting crowded systems without too much trouble, even in very complex cases (51,56-58).

If we use an inadequate physical model, we will get an inadequate result, but we cannot know what part of the model is inadequate until we compare with experiments under a range of conditions. Then, we can appropriately improve the model and move along towards our goal of understanding and controlling of biophysical systems.

This view of ionic solutions may be new to many biophysicists and so seem idiosyncratic. This view, however, has been presented to (and refereed by) communities in applied mathematics and physics (1,188), physical chemistry (97,98,176,285), physiology (99), and biophysical chemistry (98).

## <u>Acknowledgement</u>

I am indebted to the referees and editor who made suggestions that substantially improved this paper, in my view. It has been a joy to work with my many collaborators. Everything is a joint effort.

This work was made possible by the administrative work I have not been asked to do. I am grateful to the Miller Institute (University of California Berkeley), Rush University and the Bard Endowed Chair that made that possible.

# Figure Captions

Fig. 1. Control variables. **Diameter controls selectivity** of the DEKA Na channel**.** The selectivity (the ***ratio*** of the $Na^+$ and $K^+$ contents) contents of the DEKA Na channel depends on the diameter but not on the dielectric coefficient. The structural parameter (diameter) determines the selectivity. The solvation energy parameter (dielectric coefficient) determines the content. Each variable has an effect on one biological characteristic and not on the other. Thus, there are two independent ('orthogonal') control variables in the model. Loosely speaking, structure (diameter) control selectivity; solvation (dielectric) controls contents and thus conductance. This figure is redrawn from Fig. 8 of (147).

Fig. 2. Control variables: **dielectric coefficient controls contents**, i.e., conductance of the DEKA Na channel. Note the different scales on the ordinate for Na$^+$ blue and K$^+$ red. The solvation energy parameter (the dielectric coefficient) determines the contents and thus the conductance. The structural parameter (the diameter of the channel) determines the selectivity (the ratio of Na$^+$ to K$^+$). Each parameter has an effect on one biological characteristic and not on the other. Thus, there are two independent ('orthogonal') control parameters that control the biological characteristics of the model, independently. Loosely speaking, structure (diameter) control selectivity; solvation (dielectric) controls contents and thus conductance. The data shown are unexpected results of 12 simulations, each taking billions of calculations and several days. This figure is redrawn from Fig. 9 of (147).